\documentclass[%
preprint%
 ,secnumarabic%
,superscriptaddress%
,amssymb, amsmath,nobibnotes,aps]{revtex4-1}

\usepackage{epsfig}%
\usepackage{graphicx}%
\usepackage{color}
\usepackage[colorlinks=true,linkcolor=blue]{hyperref}%
\expandafter\ifx\csname package@font\endcsname\relax\else
 \expandafter\expandafter
 \expandafter\usepackage
 \expandafter\expandafter
 \expandafter{\csname package@font\endcsname}%
\fi

\begin{document}

\title{Leptophilic dark matter in gauged $L_{\mu}-L_{\tau}$ extension of MSSM}
\author{Moumita Das}
\email{moumita@prl.res.in}
\affiliation{Physical Research Laboratory, Ahmedabad 380009,
India}
\author{Subhendra Mohanty}
\email{mohanty@prl.res.in}
\affiliation{Physical Research Laboratory, Ahmedabad 380009,
India}
\def\be{\begin{equation}}
\def\ee{\end{equation}}
\def\al{\alpha}
\def\bea{\begin{eqnarray}}
\def\eea{\end{eqnarray}}


\begin{abstract}
Positron excess upto energies $\sim$350 GeV has been observed by AMS-02 result and it is consistent
with the positron excess observed by PAMELA upto 100 GeV. There is no observed excess of anti-protons
over the expected CR background. We propose a leptophilic dark matter with an $U(1)_{L_{\mu}-L_{\tau}}$
gauge extension of MSSM. The dark matter is an admixture of the $L_{\mu}-L_{\tau}$ gaugino and
fermionic partners of the extra $SU(2)$ singlet Higgs boson, which break the $L_{\mu}-L_{\tau}$
symmetry. We construct the SM$\otimes U(1)_{ L_{\mu}-L_{\tau}}$ SUSY model which provides the correct
relic density of dark matter and is consistent with constrain on $Z'$ from LHC. The large dark matter
annihilation cross-section into $\mu^{+}\mu^{-}$ and  $\tau^{+}\tau^{-}$, needed to explain PAMELA
and AMS-02 is achieved by Breit-Wigner resonance.
\end{abstract}
\maketitle

\section{Introduction}
The AMS-02 experiment reports an excess of the positron over the CR background upto
energies $\sim$350 GeV \cite{AMS-02}, consistent with the positron excess reported by PAMELA
upto 100 GeV \cite{PAMELA}. In either experiment there is no corresponding excess seen in
the anti-protons flux over the CR background. To explain the positron excess of AMS-02 and PAMELA
interms of $10^2$-$10^3$ GeV scale dark matter annihilation, would require the annihilation cross-section
of $\chi\chi\longrightarrow\mu^{+}\mu^{-},\tau^{+}\tau^{-}$ of order of $10^{-23}\,{\rm cm}^3\,{\rm s}^{-1}$
\cite{Cirelli:2008pk,DeSimone:2013fia,Yuan:2013eja}, but no annihilation into baryons or $e^{+}e^{-}$ channels (to avoid a line spectrum
in the positron signal). Dark matter with relic density consistent with PLANCK \cite{Ade:2013zuv} and WMAP
\cite{Hinshaw:2012aka} observations arises in the neutralino sector of supersymmetric extension of the
standard model. In the minimal  supersymmetric extension of the standard model, the heavy dark matter
in the TeV scale can be Winos \cite{Hisano:2006nn}. In MSSM the annihilation of Wino dark matter
is primarily into $W^{+}W^{-}$ and gives a significant anti-proton flux, which is not seen by PAMELA
or AMS-02 \cite{Cirelli:2008pk,DeSimone:2013fia,Yuan:2013eja,Adriani:2010rc}. In this paper we construct a leptophilic dark matter
by considering an $U(1)_X$
extension of MSSM, where $X=L_{\mu}-L_{\tau}$ gauge symmetry. It is known that in standard model, one
can have anomaly free gauged symmetry \cite{Foot:1991,Foot:1994vd} if the gauge charges are as following,
\begin{eqnarray}
(1)\,\, B-L\,\,,\,\,(2)\,\, L_1 =L_{e} -L_{\mu}\,\,,\,\,(3)\,\,L_2 =L_{e} -L_{-\tau}\,\,,\,\,(4)\,\,L_3 =L_{\mu} -L_{\tau}
\end{eqnarray}
Leptophilic dark matter from $U(1)_{B-L}$ extension was considered to study the baryon asymmetry \cite{Kohri:2009yn}.
We choose the $L_{\mu} -L_{\tau}$ symmetry to get $\mu^{+}\mu^{-}$ and  $\tau^{+}\tau^{-}$ in the final states
dark matter annihilation and to avoid hard positrons and anti-protons in the final states. The $L_{\mu} -L_{\tau}$
symmetry is also a natural frame work for constructing the see-saw model, which can explain the large
$\mu-\tau$ mixing in the light neutrinos \cite{Adhikary:2006rf}. If the $L_{\mu} -L_{\tau}$
gauge boson $Z'$ is very light, few GeV, then this extended model can explain the muon anomalous magnetic dipole moment
\cite{Baek:2001kca,Baek:2008nz,Heeck:2011wj}. In supersymmetry one has to introduce
two extra superfields $\hat{\eta}$ and $\hat{\bar{\eta}}$ to cancel the anomalies from the $L_{\mu} -L_{\tau}$
gauginos. the scalar components of $\hat{\eta}$ and $\hat{\bar{\eta}}$ also provide masses to the $L_{\mu} -L_{\tau}$
gauge boson $Z'$. The  mass of $Z'$, $m_{Z'}$ is greater than 2 TeV from recent LHC searches \cite{mzprime}.
There can be kinetic mixing between the field strength of the superfields \cite{Suematsu:1998wm}. This
gives the kinetic mixing between $U_{X}$ and $U_{Y}$ gauge fields and also a mixing between gauginos
$\tilde{B}$ and $\tilde{B^{\prime}}$. Since $m_{Z'}$ is much larger than $m_{Z}$, the kinetic mixing
is required to be small. This in term means that the mixing between $H_u$, $H_d$, $\eta$ and $\bar{\eta}$
in the D-term of the Higgs potential, is small. There is no significant contribution to the tree level lightest
CP even Higgs mass from the $\eta$ and $\bar{\eta}$ sector.

The dark matter is a combination of the $U_{L_{\mu}-L_{\tau}}$ gaugino, $\tilde{B^{\prime}}$ and the fermionic
partner of $\eta$ and $\bar{\eta}$. By taking $m_{Z'}\sim 2m_{\tilde{\chi}_1}$, we can make the thermal relic density
close to the PLANCK observed value of $\Omega_{\chi}h^2=0.1199\pm0.0027$ at 68$\%$ CL\cite{Ade:2013zuv}.
In the early universe the center of mass energy is not exactly coincident with the mass of $Z'$ in the propagator,
but in the present era the dark matter velocity is $\sim 10^{-3}$ and we have annihilation close to the
resonance peak \cite{Feldman:2008xs,Ibe:2008ye}. By parameterizing the dark matter mass in terms of $m_{Z'}$ as,
$ m_{Z'}^2=4m^2_{\tilde{\chi_1}}\left(1-\delta\right)$ and choosing $\delta\sim 10^{-3}$, we obtain the required
boost factor of 100 needed to explain the AMS-02 flux \cite{DeSimone:2013fia}.

The particle content of the model is described in section-\ref{sec2}, where we give the superpotential,
the kinetic mixing and neutral part of the scalar potential. In section-\ref{sec3} we discuss the mass
matrices of  CP-odd scalar, CP-even scalar and the $7\times7$ neutralino mass matrix arising from gauge eigenstates,
$\lbrace \tilde{B},\tilde{W},\tilde{H_d},\tilde{H_u},\tilde{B^{\prime}},\tilde{\eta},\tilde{\bar{\eta}}\rbrace$.
In section \ref{sec6} we give the numerical results for scalar and neutralino masses. We also
compute the thermal relic density at decoupling, the boost factor of our model in the present era with Breit-Wigner
resonance and direct detection constraints. Finally we draw the conclusion of our work in section-\ref{sec7}.
\section{model}\label{sec2}
\subsection{Particles and Superpotential}\label{sec2a}
We study the supersymmetric model of the $SU(3)_C \otimes SU(2)_L\otimes U(1)_Y \otimes U(1)_X$
gauge theory where, $X={L_{\mu}-L_{\tau}}$. The particle content of the theory is given in Table.(\ref{ptc}).
In addition to the MSSM, we need to add two fields $\hat{\eta}$ and $\hat{\bar{\eta}}$ to cancel the
anomalies from gauginos of $U(1)_X$.
The superpotential of this model can be written as,
\begin{eqnarray}
\mathcal{W}= \mu\hat{H_u}\hat{H_d}-\mu^{\prime}\hat{\eta}\hat{\bar{\eta}}+\mathcal{W}_y
\end{eqnarray}
where,
\begin{eqnarray}
 \mathcal{W}_y = Y_u \hat{U}\hat{Q}\hat{H_u}-Y_d \hat{D}\hat{Q}\hat{H_d}-Y_e \hat{E}\hat{L}\hat{H_d}
\end{eqnarray}
which gives the yukawa interaction of the fermions.
\begin{center}
\begin{table}[h]
\begin{tabular}{|c|c|c|c|c|c|c|c|c|c|c|c|c|}\hline
\begin{tabular}{c} Super Fields$\rightarrow$\\\hline Groups$\downarrow$\end{tabular}
& $H_u$ & $H_d$ & $L_{\tau}$ & $L_{\mu}$ & $L_{e}$ & $\, \eta\, $ & $\, \bar{\eta}\, $  \\\hline
$SU(3)$ & 1 & 1 & 1 & 1 & 1  & 1 & 1 \\\hline
$SU(2)$ & 2 & 2 & 2 & 2 & 2  & 1 & 1 \\\hline
$U(1)_Y $& $\frac{1}{2}$ &  $-\frac{1}{2}$ & $-\frac{1}{2}$  &  $-\frac{1}{2}$ & $-\frac{1}{2}$   & 0 & 0 \\\hline
$U(1)_{x}$ & 0 & 0 & $-\frac{1}{2}$ &  $\frac{1}{2}$ & 0 & -1 & 1  \\\hline
\end{tabular}
\caption{Particle content of the theory}
\label{ptc}
\end{table}
\end{center}
In this model, there will be an extra gaugino corresponding to the extra $U(1)_X$ group.
The extra symmetry breaking part
due to the presence of $L_{\mu}-L_{\tau}$ symmetry are as follows,
\begin{eqnarray}
\mathcal{L}_{SB}&=&M_{B B^{\prime}}\tilde{B}\tilde{B}^{\prime}-\frac{M_{B^{\prime}}}{2}\tilde{B}^{\prime}\tilde{B}^{\prime}
-m^2_{\eta}|\eta^2|-m^2_{\bar{\eta}}|\bar{\eta}^2|-\eta\bar{\eta}B_{\mu^{\prime}}
\label{SB}
\end{eqnarray}

\subsection{Gauge Mixing}\label{sec2b}

In this model, there is a gauge kinetic mixing between $U(1)_X$ and $U(1)_Y$ gauge groups.
The general description of gauge kinetic mixing in supersymmetry is discussed in \cite{Suematsu:1998wm}.
In Wess-Zumino gauge, the relevant part of the Lagrangian is as follows,
\begin{eqnarray}
\mathcal{L}=\left.\frac{1}{32}W^{\alpha}W_{\alpha}\right|_F+\left.\Phi^{\dagger} e^{2gQV}\Phi\right|_D
\end{eqnarray}
For two different $U(1)$ gauge group, there will be two different $W_{WZ}$ along with their gauge field,
gaugino and the auxillary field. Therefore the gauge kinetic term will be
\begin{eqnarray}
 \sim W^{a\delta}W^{a}_{\delta} + W^{b\delta}W^{b}_{\delta} + W^{a\delta}W^{b}_{\delta}
\end{eqnarray}
Choosing the appropriate basis, we can eliminate the mixing term, but the effect of that is reflected in the
gauge covariant derivative as,
\begin{eqnarray}
D_{\mu}=\partial_{\mu}-ig_a q_a A^{\mu}_a-i\left(g_{ab}q_a+g_b q_b\right)A^{\mu}_b
\label{g.k.m.}
\end{eqnarray}
Similarly gauginos $\hat{\lambda}_{a,b}$ and auxillary fields $\hat{D}_{a,b}$ in supersymmetry
also change as,
\begin{eqnarray}
\hat{g}_a q_a \hat{\lambda}^a +\hat{g}_b q_b \hat{\lambda}^b=g_a q_a\lambda^a+\left(g_{ab}q_a+g_b q_b\right)\lambda^b
\label{g.k.m.}
\end{eqnarray}
and similar expression is valid for auxillary fields. 
Due to the mixing, the auxillary fields $D_{a,b}$ can be written as,
\begin{eqnarray}
D_a =-\sum_i g_a\,q_a^i |\phi_i|^2 \quad,\quad D_b =-\sum_i \left( g_{ab}\,q_a^i +  g_b\,q_a^i\right)|\phi_i|^2
\end{eqnarray}
In our model, $g_a$, $g_b$ and $g_{ab}$ describe the $U(1)_{Y}$,  $U(1)_{X}$ gauge coupling constant, $g_1$, $g_X$ and the
mixing $g_m$. This kinetic mixing modifies the D-term as,
\begin{eqnarray}
V_{D}|_{U(1)}=\frac{1}{2}D_a^2 +\frac{1}{2}D_b^2
\label{DU1}
\end{eqnarray}
Eq.~(\ref{DU1}) denotes the only contribution from two $U(1)$ groups. This mixing effects the Higgs sector as
well as neutralino sector through the change in the auxillary fields. There will be contribution in D-term from
$SU(2)$, similar to the MSSM and the total D-term for our model is,
\begin{eqnarray}
V_D=\left.\frac{g_1^2+g_2^2}{8}\right.\cdot\frac{\left(H_u^2+H_d^2\right)^2}{4}
+\frac{1}{8}\left[\frac{1}{2}g_m\left(H_u^2-H_d^2\right)-\frac{1}{2}g_X\left(\eta^2-\bar{\eta}^2\right)\right]^2
\label{Dtotal}
\end{eqnarray}

\subsection{Neutral Part of Scalar Potential}\label{sec2c}
Neutral part of scalar potential contain three parts as,
\begin{eqnarray}
V=V_{SB}+V_F+V_D
\end{eqnarray}
The soft symmetry breaking part of the neutral scalar is given by,
\begin{eqnarray}
V_{SB}=m_u^2|H_u|^2+m_d^2|H_d|^2-B_{\mu}H_u H_d+m_{\eta}^2|\eta|^2
+m_{\bar{\eta}}^2|\bar{\eta}|^2-B_{\mu^{\prime}}\eta\bar{\eta}
\end{eqnarray}
where the masses $m_i,\, (i=u,d,\eta,\bar{\eta})$ can be replaced using the minimization conditions,
discussed below.

The F-term of the superpotential can be derived as,
\begin{eqnarray}
V_{F}=\left|\frac{\partial\mathcal{W}(\phi)}{\partial\phi}\right|^2
\end{eqnarray}
where $\phi$ denotes the corresponding scalar field of each superfields. Hence the F-term becomes,
\begin{eqnarray}
V_{F}= \frac{\mu^2}{2}\left(H_u^2+H_d^2\right)+\frac{\mu^{\prime2}}{2}\left(\eta^2+\bar{\eta}^2\right)
\end{eqnarray}
In the next section we will present the minimization condition of the scalar potential and derive the
CP-even, CP-odd and neutralino mass matrices.
\section{mass Spectrum}\label{sec3}
\subsection{Minimization conditions}
The extended gauge symmetry $SU(3)_C \otimes SU(2)_L\otimes U(1)_Y \otimes U(1)_{L_{\mu}-L_{\tau}}$ breaks into
$SU(3)_C \otimes SU(2)_L\otimes U(1)_{em}$ when the Higgs doublets $H_u$ and $H_d$ and the other two scalars
$\eta$ and $\bar{\eta}$ acquire vevs as,
\begin{eqnarray}
\langle H_u \rangle = v_u /\sqrt{2}\quad,\quad\langle H_d \rangle =v_d/\sqrt{2}\quad,\quad
\langle \eta \rangle =v_{\eta}/\sqrt{2}\quad\&\quad\langle \bar{\eta} \rangle = v_{\bar{\eta}}/\sqrt{2}.
\end{eqnarray}
The neutral part of the scalar fields can be written in terms of real and imaginary part as follows,
\begin{eqnarray}
&H_u& = \frac{1}{\sqrt{2}}\left(H_{uR}+v_u+i H_{uI}\right)\quad\&\quad H_d = \frac{1}{\sqrt{2}}\left(H_{dR}+v_d+i H_{dI}\right)\\
&\eta& = \frac{1}{\sqrt{2}}\left(\eta_{R}+v_\eta+i \eta_{I}\right)\quad\&\quad
\bar{\eta} = \frac{1}{\sqrt{2}}\left(\bar{\eta}_{R}+v_{\bar{\eta}}+i \bar{\eta}_{I}\right)
\end{eqnarray}
We parametrize the vev of Higgs doublet and extra singlets as follows,
\begin{eqnarray}
tan\beta=\frac{v_u}{v_d} \quad, \quad v_u^2+v_d^2=v^2\quad, \quad
tan\beta^{\prime}=\frac{v_{\eta}}{v_{\bar{\eta}}} \quad, \quad v_{\eta}^2+v_{\bar{\eta}}^2=v_0^2;
\end{eqnarray}

Now, the minimization conditions  at tree level,
$\frac{\partial V}{\partial H_{uR} }=\frac{\partial V}{\partial H_{dR} }=
\frac{\partial V}{\partial \eta_{R} }=\frac{\partial V}{\partial \bar{\eta}_{R} }=0$
can be written as,
\begin{subequations}
\begin{eqnarray}
m_u^2&=&B_{\mu}tan\beta-|\mu^2|-\frac{1}{8}\left(g_1^2+g_2^2+g_m^2\right)(v_d^2-v_u^2)-\frac{1}{4}g_m g_X\left(v_{\eta}^2-v_{\bar{\eta}}^2\right)\\
m_d^2&=&B_{\mu}cot\beta-|\mu^2|+\frac{1}{8}\left(g_1^2+g_2^2+g_m^2\right)(v_d^2-v_u^2)+\frac{1}{4}g_m g_X\left(v_{\eta}^2-v_{\bar{\eta}}^2\right)\\
m_{\eta}^2&=&B_{\mu^{\prime}}cot\beta^{\prime}-|\mu^{\prime 2}|-\frac{1}{4}g_m g_X(v_d^2-v_u^2)-\frac{1}{2} g_X^2\left(v_{\eta}^2-v_{\bar{\eta}}^2\right)\\
m_{\bar{\eta}}^2&=&B_{\mu^{\prime}}tan\beta^{\prime}-|\mu^{\prime 2}|+\frac{1}{4}g_m g_X(v_d^2-v_u^2)+\frac{1}{2} g_X^2\left(v_{\eta}^2-v_{\bar{\eta}}^2\right)
\label{mcond}
\end{eqnarray}
\end{subequations}
These relations are used to replace the soft-breaking parameters in the scalar mass matrices.
\subsection{CP-odd Scalar:}\label{4CPO}
The CP-odd mass matrix contain the imaginary part of the neutral scalar and the basis is
$\lbrace H_{dI},H_{uI},\eta_I,\bar{\eta}_I \rbrace$.
At tree level, Higgs doublet and extra singlet do not mix and the pseudo-scalar mass matrix becomes,
\begin{subequations}
\begin{eqnarray}
Mp_{11}^2&=& B_{\mu}tan\beta\\
Mp_{22}^2&=& B_{\mu}cot\beta\\
Mp_{12}^2&=&Mp_{21}^2=B_{\mu} \\
Mp_{33}^2&=& B_{\mu^{\prime}}tan\beta^{\prime}\\
Mp_{44}^2&=& B_{\mu^{\prime}}cot\beta^{\prime}\\
Mp_{34}^2&=&Mp_{43}^2=B_{\mu^{\prime}}
\end{eqnarray}
\end{subequations}
and remaining 8 components of $4\times4$ matrix are zero. The two CP-odd massive scalars have masses,
\begin{eqnarray}
m^2_A=\frac{2 B_{\mu}}{sin2\beta} \quad\&\quad m^2_{A\eta}=\frac{2 B_{\mu^{\prime}}}{sin2\beta^{\prime}}
\end{eqnarray}
and the other two CP-odd eigen-vectors represent the massless Goldstone bosons, provide longitudinal
components of $Z$ and $Z'$ gauge bosons. There is no mixing between $\lbrace H_{dI},H_{uI}\rbrace$ and
$\lbrace\eta_I,\bar{\eta}_I \rbrace$ sectors.
\subsection{CP-Even Scalar:}\label{4CPE}
The basis of the CP-even mass matrix is considered as $\lbrace H_{dR},H_{uR},\eta_R,\bar{\eta}_R \rbrace$ and in this basis,
the mass matrix has 10 independent entities as follows;
\begin{subequations}
\begin{eqnarray}
Ms_{11}^2&=&B_{\mu}tan\beta+\frac{1}{4}\left(g_1^2+g_2^2+g_m^2\right)v_d^2=m^2_A sin^2\beta+\frac{1}{4}\left(g_1^2+g_2^2+g_m^2\right)v_d^2\\
Ms_{22}^2&=&B_{\mu}cot\beta+\frac{1}{4}\left(g_1^2+g_2^2+g_m^2\right)v_u^2=m^2_A cos^2\beta+\frac{1}{4}\left(g_1^2+g_2^2+g_m^2\right)v_u^2\\
Ms_{33}^2&=&B_{\mu^{\prime}}cot\beta^{\prime}+g_X^2 v_{\eta}^2=m^2_{A\eta} cos^2\beta^{\prime}+g_X^2 v_{\eta}^2\\
Ms_{44}^2&=&B_{\mu^{\prime}}tan\beta^{\prime}+g_X^2 v_{\bar{\eta}}^2=m^2_{A\eta} sin^2\beta^{\prime}+g_X^2 v_{\bar{\eta}}^2\\
Ms_{12}^2&=&-B_{\mu}-\left(g_1^2+g_2^2+g_m^2\right)v_u v_d =-m^2_A sin\beta cos\beta-\left(g_1^2+g_2^2+g_m^2\right)v_u v_d\\
Ms_{13}^2&=&\frac{1}{2}g_m g_X v_d v_{\eta}\\
Ms_{14}^2&=&-\frac{1}{2}g_m g_X v_d v_{\bar{\eta}}\\
Ms_{23}^2&=&-\frac{1}{2}g_m g_X v_u v_{\eta}\\
Ms_{24}^2&=&\frac{1}{2}g_m g_X v_u v_{\bar{\eta}}\\
Ms_{34}^2&=&-B_{\mu^{\prime}}-g_X^2 v_{\eta} v_{\bar{\eta}}=-m^2_{A\eta} sin\beta^{\prime} cos\beta^{\prime}-g_X^2 v_{\eta} v_{\bar{\eta}}
\end{eqnarray}
\end{subequations}
We will numerically diagonalize the mass matrix and identify the lightest CP-even mass as the 125 GeV Higgs (after
radiative corrections).
\subsection{Neutralino:}\label{4neu}
In this model, there are three extra neutralinos corresponding to the extended gauge symmetry. The following neutralino-mass matrix is written
in terms of the basis
$\lbrace \tilde{B},\tilde{W},\tilde{H_d},\tilde{H_u},\tilde{B^{\prime}},\tilde{\eta},\tilde{\bar{\eta}}\rbrace$,
\vspace*{0.5cm}
\begin{eqnarray}
M_{neu}=\left(\begin{array}{ccccccc}
M_1& 0 & -\frac{1}{2}g_1 v_d & \frac{1}{2}g_1 v_u & \frac{1}{2}M_{BB^{\prime}}& 0 & 0 \\
0& M_2 & \frac{1}{2}g_2 v_d & -\frac{1}{2}g_2 v_u & 0& 0 & 0 \\
-\frac{1}{2}g_1 v_d & \frac{1}{2}g_2 v_d & 0& -\mu & -\frac{1}{2}g_m v_d& 0 & 0 \\
\frac{1}{2}g_1 v_u & -\frac{1}{2}g_2 v_u & -\mu & 0 & \frac{1}{2}g_m v_u & 0 & 0 \\
\frac{1}{2}M_{BB^{\prime}}& 0 & -\frac{1}{2}g_m v_0& \frac{1}{2}g_m v_u & M_{B} & -g_X v_{\eta} & -g_X v_{\bar{\eta}} \\
0 & 0 & 0& 0 & -g_X v_{\eta}& 0 & -\mu^{\prime}\\
0& 0 & 0& 0 & -g_X v_{\bar{\eta}}& -\mu^{\prime} & 0
\end{array} \right),
\end{eqnarray}
\\
In general, neutralino LSP will be the mixture of seven gauge eigen states
$\lbrace \tilde{B},\tilde{W},\tilde{H_d},\tilde{H_u},\tilde{B^{\prime}},\tilde{\eta},\tilde{\bar{\eta}}\rbrace$.
However appropriate choice of parameter $\lbrace M_1,M_2,M_{B'},\mu,\mu'\rbrace$ define the characteristics
of the neutralino LSP \cite{Basso:2012gz,O'Leary:2011yq}.
\section{Numerical analysis and Results}\label{sec6}
The result of this model can be studied in term of various free parameters. Among them,
$\lbrace g_1,g_2,g_x,g_m,v,tan\beta,v_0,tan\beta^{\prime} \rbrace$ are gauge parameters and
$\lbrace B_{\mu},B_{\mu^{\prime}},m_u^2,m_d^2,m_{\eta}^2,m_{\bar{\eta}}^2\rbrace$ are the soft breaking parameters.
We have used the well measured value for  $U(1)_Y$ and $SU(2)$ gauge coupling $g_1$ and $g_2$ respectively. The
vev of the extra singlets determines the mass of the $Z^{\prime}$, $m_{Z^{\prime}}$ of the extra $U(1)_X$.
The masses
$\lbrace m_u^2,m_d^2,m_{\eta}^2,m_{\bar{\eta}}^2\rbrace$ are replaced interm of $\lbrace\mu$, $\mu^{\prime}$,
$B_{\mu}$, $B_{\mu^{\prime}}$, $v$, $tan\beta$, $v_0$, $tan\beta^{\prime}\rbrace$ using the minimization conditions
given in Eq.~(\ref{mcond}). Hence the independent parameters to calculate the mass spectrum is as follows,
\begin{eqnarray}
\left(g_x,g_m,tan\beta,tan\beta^{\prime},m_{Z^{\prime}},\mu,\mu^{\prime},B_{\mu},B_{\mu^{\prime}}\right)
\end{eqnarray}
We have taken $\mu,\mu^{\prime}\simeq1$ TeV and the other parameters are as
follows,

\begin{center}
\begin{tabular}{|c|c|c|c|c|}\hline
$\,\, g_x\,\, $ & $\,\,g_m\,\,$ & $\,tan\beta\,$ & $\,tan\beta^{\prime}\,$ &$\,m_{Z^{\prime}}\,$\\\hline
0.2 & 0.01 & 40 & 10.2&  2143.6 TeV\\\hline
\end{tabular}
\end{center}
\subsection{Masses for CP-even Scalar:}
Using the above mentioned values of the parameters at electroweak scale, we can find the following masses
for the CP-even scalars.
\begin{center}
\begin{table}[ht]
\begin{tabular}{|c|c|c|c|c||}\hline
$m_{h,1}$ & $m_{h,2}$ & $m_{h,3}$ & $m_{h,4}$\\\hline
91.3 & 99.5 & 200.1 & 2144.8\\\hline
\end{tabular}
\caption{Masses are in GeV}
\label{CP_MASS1}
\end{table}
\end{center}
In this model, we have considered moderate value for kinetic mixing between the two $U(1)$ gauge group. However
this mixing has no significant effect in raising the tree level mass of lowest Higgs beyond what is there in MSSM.
Hence we required loop correction to increase the value of lowest Higgs mass and choosing stop mass around $\sim 1$ TeV,
we get correct enhancement.
\begin{eqnarray}
m_{h,1}=\sqrt{91.3^2+\frac{3 g_2^2}{16\pi^2 M_w^2}\,\frac{M_t^2}{{\rm sin}^2 \beta}\,
Log\left(\frac{m^2_{\tilde{t_1}}\,m^2_{\tilde{t_2}}}{M_t^4}\right)}\simeq 125 {\rm GeV}
\end{eqnarray}
where $M_{\rm w}$ and $M_{\rm t}$ denote the W-boson and top quark mass and $m_{\tilde{t_i}}$ is the mass of the stop.
\subsection{Masses for CP-odd Scalar and Charged Scalar:}
The masses for CP-odd Scalar are given by,
\begin{eqnarray}
m^2_A=\frac{2 B_{\mu}}{{\rm sin} 2\beta}=\left(200.1\,\, {\rm GeV}\right)^2 \quad\&
\quad m^2_{A\eta}=\frac{2 B_{\mu^{\prime}}}{{\rm sin}2\beta^{\prime}}=\left(101.5\,\, \rm{GeV}\right)^2
\end{eqnarray}
The mass of the charged scalar is exactly similar to MSSM at tree level and it is,
\begin{eqnarray}
m^{2}_{h^{+}}=m^2_A+m^2_w =\left(219.8 \,{\rm GeV}\right)^2
\end{eqnarray}
\subsection{Neutralino}
To calculate mass of the neutralinos, we have to fix other free parameters, gaugino masses
$M_1,M_2,M_{B}\,{\rm and}\,M_{BB^{\prime}}$. Gauge kinetic mixing also give the mixing between MSSM neutralino sector
and the extra neutralinos $\tilde{B^{\prime}},\tilde{\eta}$ and $\tilde{\bar{\eta}}$ coming from $U(1)_X$.
If we choose the parameters such that $\mu^{\prime}\ll\left( M_1,M_2,M_{B^{\prime}},\mu\right)$,
then we will get the lightest supersymmetric particle (LSP) as a mixture of
$\tilde{\eta}\,\,{\rm and}\,\,\tilde{\bar{\eta}}$. Hence the chosen parameters and the mass
of the neutrinos are given in Table~\ref{Neu_MASS}.
\begin{center}
\begin{table}[ht]
\begin{tabular}{|c|c|c|c||c|c|c|c|c|c|c|}\hline
$\,M_1\,$ & $\,M_2\,$ & $\,M_{B^{\prime}}\,$ & $\,M_{BB^{\prime}}\,$&$\,m_{\tilde{\chi_1}}\,$ & $\,m_{\tilde{\chi_2}}\,$
& $\,m_{\tilde{\chi_3}}\,$
& $\,m_{\tilde{\chi_4}}\,$& $\,m_{\tilde{\chi_5}}\,$& $\,m_{\tilde{\chi_6}}\,$& $\,m_{\tilde{\chi_7}}\,$\\\hline
4000 & 4000 & 4420 & 1000& 1072.34 & 1971.7 & 2995.98 & 3000.63&3865.87 & 4003.55& 5454.6 \\\hline
\end{tabular}
\caption{Soft breaking parameters and neutralino masses in GeV}
\label{Neu_MASS}
\end{table}
\end{center}
\subsubsection{{\textbf{Lightest supersymmetric particle (LSP):}}}
In this model, lightest supersymmetric particle, which we identify as the dark matter,
has the mass of $1072.6$ GeV and in terms of gauge eigenstates it is,
\begin{eqnarray}
\tilde{\chi_1}= -0.05\,\tilde{B}-5.94\times 10^{-6}\,\tilde{W}-6.99\times 10^{-4}\,\tilde{H_d}-2.35\times 10^{-4}\,\tilde{H_u}
+\nonumber\\
0.294\,\tilde{B^{\prime}}+0.527\,\tilde{\eta}+0.795\,\tilde{\bar{\eta}}
\end{eqnarray}
This dark matter consisting of standard model singlets $\tilde{\eta}$ and $\tilde{\bar{\eta}}$ has small annihilation
channels into standard model particles. So in general their relic density is large similar to that is seen
in other gauge extension \cite{O'Leary:2011yq,Basso:2012gz}. Therefore we consider a resonance channel through
$Z^{\prime}$ by taking $2m_{\tilde{\chi_1}}=m_{Z^{\prime}}$ to get correct relic density. In our model the dominant
channel for getting correct relic abundance is,
$\,\,\tilde{\chi_1}\tilde{\chi_1}\rightarrow Z^{\prime}\rightarrow \tau^{+}\tau^{-}\,,\,\mu^{+}\mu^{-}$. The cross section
of the resonance channel is velocity dependent and hence it can simultaneously satisfy both the
relic abundance at freeze-out and positron excess of DM signal \cite{Feldman:2008xs,Ibe:2008ye,Guo:2009aj,Bi:2009uj}.
Thermal average of cross-section is calculated at freeze-out, $x=m/T\approx20$ and hence velocity,
$v\approx\sqrt{3/x}$, whereas the dark matter annihilation occurs at $x\approx3\times10^{6}$ or
$v\approx10^{-3}$. To explain the positron excess of AMS-02 results, we require the cross section much
larger than what we need for relic density and this require boost factor of $\mathcal{O}(100-1000)$
for cross-section.
\subsubsection{{\textbf{Breit-Wigner resonance and boost factor:}}}
The annihilation cross-section for the dominant channel will be,
\begin{eqnarray}
\sigma v=\frac{\bar{g}_1^2 }{4\pi}\frac{m^2_{\tilde{\chi_1}}}{\left(s-m_{Z'}^2\right)^2+\Gamma_{Z'}^2 m_{Z'}^2}
\label{sigmav1}
\end{eqnarray}
where the coupling $\bar{g}_1\simeq0.07$ depends on the coupling constant of extra $U(1)$ symmetry, the mixing
between two $U(1)$ gauge groups and fraction of fermionic partner of $\eta$ and $\bar{\eta}$ in the
LSP, $\tilde{\chi_1}$. The decay width of the $Z'$ boson signifies by $\Gamma_{Z'}$, which
has dominant contribution from the decay mode, $Z^{\prime}\rightarrow \tau^{+}\tau^{-}\,,\,\mu^{+}\mu^{-}$ and it is given by,
\begin{eqnarray}
\Gamma_{Z'}=\frac{\bar{g}_2^2 }{4\pi}\,m_{Z'}
\end{eqnarray}
Here the coupling, $\bar{g}_2\simeq0.1$ depends on the coupling constant of extra $U(1)$ symmetry and the mixing
between two $U(1)$ gauge groups.
Using the non-relativistic limit of $s$, $s=4m^2_{\tilde{\chi_1}}\left(1+v^2/4\right)$,
Eq.~(\ref{sigmav1}) can be simplified into,
\begin{eqnarray}
\sigma v=\frac{\bar{g}_1^2 }{64\pi m_{\tilde{\chi_1}^2}}\frac{1}{\left(\delta+v^2/4\right)^2+\gamma^2 }
\label{sigmav2}
\end{eqnarray}
where $ m_{Z'}^2=4m^2_{\tilde{\chi_1}}\left(1-\delta\right)$ and
$\gamma^2=\Gamma_{Z'}^2\left(1-\delta\right)/m^2_{\tilde{\chi_1}}$.

Now the thermal average of annihilation rate will be as follows \cite{Feldman:2008xs,Ibe:2008ye,Guo:2009aj,Bi:2009uj},
\begin{eqnarray}
 \langle\sigma v\rangle= \frac{1}{n^2_{EQ}}\frac{m_{\tilde{\chi_1}}}{64\pi^4 x}
 \int_{4m_{\tilde{\chi_1}}}^{\infty}\hat{\sigma}(s)\sqrt{s}K_1\left(\frac{x\sqrt{s}}{m_{\tilde{\chi_1}}}\right)ds
 \label{t_avg1}
\end{eqnarray}
where,
\begin{eqnarray}\label{nEQ}
n_{EQ}&=&\frac{g_i}{2\pi^2}\frac{m^3_{\tilde{\chi_1}}}{x}K_2(x)\\
\hat{\sigma}(s)&=&2g_i^2 m_{\tilde{\chi_1}}\sqrt{s-4m^2_{\tilde{\chi_1}}}\,\sigma v
\label{hatsigma}
\end{eqnarray}
where $x\equiv m_{\tilde{\chi_1}}/T$, $K_1(x)$ and $K_2(x)$ denote the second type of modified Bessel functions
and $g_i$ is the internal degrees of freedom of dark matter.
Inserting Eq.~(\ref{nEQ}) and Eq.~(\ref{hatsigma}), in the thermally average cross-section Eq.~(\ref{t_avg1}),
it can be simplified as,
\begin{eqnarray}
 \langle\sigma v\rangle= \frac{\bar{g}_1^4}{256\pi^{3/2}}\frac{x^{3/2}}{m^2_{\tilde{\chi_1}}}
 \int_{0}^{\infty}\frac{\sqrt{z}\,{\rm Exp}[-xz/4]}{\left(\delta+v^2/4\right)^2+\gamma^2}dz
 \label{t_avg}
\end{eqnarray}
where $z$ is defined as, $v^2=z$.
At freeze-out, $x=20$, we get the correct
relic density $\Omega_{DM}h^2=0.1$, consistent with Planck results \cite{Ade:2013zuv} and the 9-year
WMAP data \cite{Hinshaw:2012aka}.
\begin{figure}[ht]
\begin{center}
\includegraphics[width=0.6\textwidth]{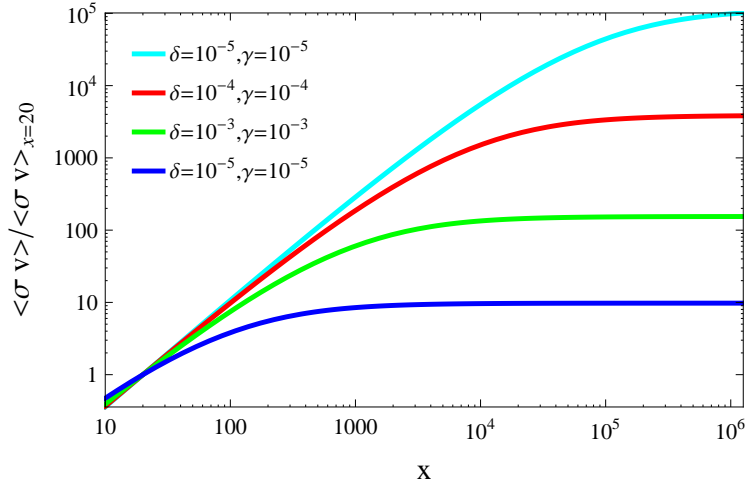}
\end{center}
\caption{\textit{Breit-Wigner enhanced relative cross-section $\rm\langle \sigma v\rangle/\langle \sigma v\rangle_{x=20}$ as a functions
 of $\rm x=m_{\tilde{\chi_1}}/T$}}
\label{diff_mhiggs}
\end{figure}
The boost factor is defined as the ratio, $\rm\langle \sigma v\rangle/\langle \sigma v\rangle_{x=20}$ is plotted in Fig-1.
This figure shows the variation of the thermally average cross-section with $x$ for different set
of values of $\delta$ and $\gamma$. We see that $\delta=10^{-3}$ and $\gamma=10^{-3}$ can give the required boost
factor of 100 to explain the AMS-02 flux of positrons \cite{DeSimone:2013fia}.
The chosen values of the parameters of our model in electroweak scale
ensure these values of $\delta$ and $\gamma$ and hence the boost factor.
\begin{figure}[h!]
\begin{center}
\includegraphics[width=0.65\textwidth]{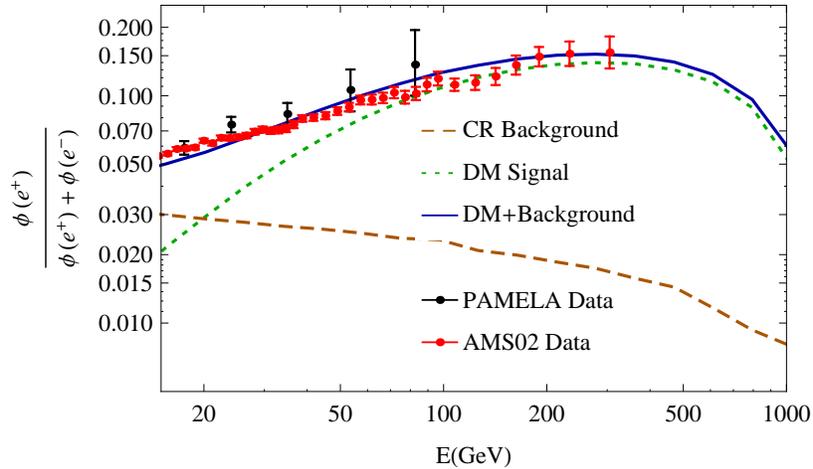}
\end{center}
\caption{\textit{The positron flux spectrum compared with data from AMS-02 \cite{AMS-02}}}
\label{pos_var}
\end{figure}

\subsubsection{{\textbf{Comparison with AMS02, PAMELA and FermiLat data:}}}

\begin{figure}[h!]
\begin{center}
\includegraphics[width=0.65\textwidth]{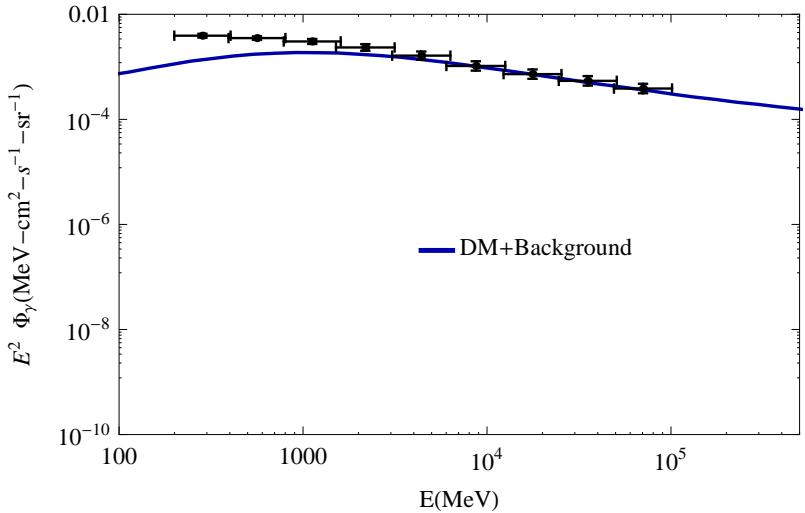}
\end{center}
\caption{\textit{The gamma ray spectrum compared with data from Fermi Lat \cite{FermiLAT:2011ab}}}
\label{gamma_var}
\end{figure}

We have calculated the positron flux coming from leptophilic dark matter to interpret the
AMS02 results using publicly available code PPPC4DMID \cite{Cirelli:2010xx,Ciafaloni:2010ti}
and GALPROP \cite{galprop}. In PPPC4DMID \cite{Cirelli:2010xx,Ciafaloni:2010ti} code
the main input is the dark matter mass of $1072.6$ GeV along with the analytical formula
given in Eq.~(\ref{t_avg1}) for the  dominant annihilation channel.
The spectral shape of positrons and gamma-ray  produced in the dark matter annihilation is calculated using PPPC4DMID. 
Then it is used as the input in GALPROP \cite{galprop} to calculate the positron and $\gamma$-ray flux which would be 
seen in experiments. In GALPROP, we have considered Burkert profile
\cite{burkert} for dark matter halo (as that requires the lowest boost factor)with the profiles,
\begin{eqnarray}
 \rho(r)=\rho_s [(1+r/r_s)(1+(r/r_s)^2)]^{-1};\,r_s=12.67\,\rm{kpc};\,\rho_s=0.712\, \rm{GeV}/\rm{cm}^3.
\end{eqnarray}
Fig.~(\ref{pos_var}) shows the dark matter interpretation of
the excess in AMS-02 positron fraction data. 
We also check the $\gamma$-rays observation for this dark matter candidate and found that
this is consistent with the experimental results \cite{FermiLAT:2011ab} as shown in Fig.(\ref{gamma_var}).

\subsubsection{{\textbf{Direct detection constraints:}}}
The relevant part of the Lagrangian for spin-independent interaction,
\begin{eqnarray}
\mathcal{L}=a_q\bar{\tilde{\chi_1}}\tilde{\chi_1}\bar{q}q
\end{eqnarray}
where coupling between neutralino and quark is denoted by $a_q$. The cross-section of dark matter
scattering off target nucleus is as follows,
\begin{eqnarray}
\sigma = \frac{4m_r^2}{\pi}\left(Z\,f_p+(A-Z)\,f_n\right)^2
\end{eqnarray}
where the reduced mass of nucleon is $m_r$. The coupling between neutralino and proton or neutron, $f_{p,n}$ is
\cite{Jungman:1995df,Bertone:2004pz},
\begin{eqnarray}
f_{p,n}=\sum_{q=u,d,s}f^{(p,n)}_{Tq}a_q\frac{m_{p,n}}{m_q}+\frac{2}{27}f^{(p,n)}_{TG}\sum_{q=c,b,t}a_q\frac{m_{p,n}}{m_q}
\end{eqnarray}
where $f^{(p,n)}_{TG}=1-\sum_{q=u,d,s}f^{(p,n)}_{Tq}$ and the values of $f^{(p,n)}_{Tq}$ are given in \cite{Ellis:2000ds}.
The coupling $a_q$ can be approximated as, $a_q/m_q=\left(g_{\tilde{\chi_1} \tilde{\chi_1} h_i}\,g_{h_i qq}\right)/\left(s-m_h^2\right)$.
In our model, the neutralino-Higgs coupling $g_{\tilde{\chi_i} \tilde{\chi_j} h_k}$ has the form of,
\begin{eqnarray}
g_{\tilde{\chi_i} \tilde{\chi_j} h_k}\simeq\sqrt{2}\left(g_1 Z^N_{i,1}-g_2 Z^N_{i,2}+g_m Z^N_{i,5}\right)
\left(Z^N_{j,3}Z^H_{k,3}-Z^N_{j,4}Z^H_{k,2}\right)\nonumber\\
+\sqrt{2}g_x\left(Z^N_{i,6}Z^N_{j,5}Z^H_{k,3}-Z^N_{i,7}Z^N_{j,5}Z^H_{k,4}\right)
\end{eqnarray}
where $Z^N_{i,j}$ and $Z^H_{i,j}$ denote the element of diagonalization matrix from gauge eigen state to mass eigen state
for neutralino and CP-even Higgs. Here $g_{h_i qq}$ is the product of the Higgs-quark coupling of MSSM and $Z^H_{i,j}$.
If the lightest CP-even Higgs is the propagator, then neutralino-Higgs coupling is highly suppressed and hence
$g_{\tilde{\chi_1} \tilde{\chi_1} h_1}\,g_{h_1 qq}\sim10^{-6}\,\,{\rm GeV}^{-1}$ and cross-section becomes $10^{-47}\,\,{\rm cm}^2$.
If the propagator is the heavy CP-even Higgs, the cross-section is even smaller. Because the Higgs-quark
coupling is  suppressed as the Heavy Higgs has dominant fraction from $\eta$ and $\bar{\eta}$ only and less from
$H_u$ and $H_d$. In this case, $g_{\tilde{\chi_1} \tilde{\chi_1} h_4}\,g_{h_4 qq}$ will be $\sim10^{-7}\,\,{\rm GeV}^{-1}$.
Hence cross-section remain well below the exclusion limit of XENON100 \cite{Aprile:2012nq}.

\section{Conclusions}\label{sec7}
We provide a SUSY model of leptophilic dark matter to explain the excess of positrons seen at AMS-02 and PAMELA.
In addition there is no antiproton excess over the CR background. We consider $L_{\mu}-L_{\tau}$ extension of MSSM
where the dark matter is a combination of $U(1)_X$ gaugino and the corresponding Higgsinos. By choosing the dark matter
mass $2m_{\tilde{\chi_1}}\simeq m_{Z^{\prime}}$, we get a resonance enhancement of the annihilation cross-section
$\,\,\tilde{\chi_1}\tilde{\chi_1}\rightarrow \tau^{+}\tau^{-}\,,\,\mu^{+}\mu^{-}$ in the present era correspond to the
freeze-out cross-section. This model provides the most economical extension of SUSY model to explain the positrons
excess seen in AMS-02 and PAMELA, consistent with other constraints from direct and indirect dark matter observations.


\end{document}